\documentclass{article}
\usepackage{spconf,amsmath,graphicx}

\usepackage{booktabs}
\usepackage{arydshln}
\usepackage{caption}
\usepackage{subcaption}
\usepackage{hyperref}
\usepackage{url}
\usepackage{multirow}
 
\title{PromptTTS: Controllable Text-to-Speech with Text Descriptions}
\name{Zhifang Guo$^{\ast}$\thanks{$\ast$ Equal contribution.}, Yichong Leng$^{\ast}$, Yihan Wu, Sheng Zhao, Xu Tan$^{\dag}$\thanks{$\dag$ Corresponding Author: Xu Tan, xuta@microsoft.com}}
\address{Univerisity of Science and Technology of China, Renmin University of China, \\ 
Microsoft Azure Speech, Microsoft Research Asia\\
\texttt{zhifangguo9@gmail.com, lyc123go@mail.ustc.edu.cn, yihanwu@ruc.edu.cn, } \\ \texttt{szhao@microsoft.com, xuta@microsoft.com}
}
\begin{document}
%
\maketitle
\begin{abstract}

Using a text description as prompt to guide the generation of text or images (e.g., GPT-3 or DALLE-2) has drawn wide attention recently. Beyond text and image generation, in this work, we explore the possibility of utilizing text descriptions to guide speech synthesis. Thus, we develop a text-to-speech (TTS) system (dubbed as PromptTTS) that takes a prompt with both style and content descriptions as input to synthesize the corresponding speech. Specifically, PromptTTS consists of a style encoder and a content encoder to extract the corresponding representations from the prompt, and a speech decoder to synthesize speech according to the extracted style and content representations. Compared with previous works in controllable TTS that require users to have acoustic knowledge to understand style factors such as prosody and pitch, PromptTTS is more user-friendly since text descriptions are a more natural way to express speech style (e.g., ``A lady whispers to her friend slowly''). Given that there is no TTS dataset with prompts, to benchmark the task of PromptTTS, we construct and release a dataset containing prompts with style and content information and the corresponding speech. Experiments show that PromptTTS can generate speech with precise style control and high speech quality. Audio samples and our dataset are publicly available\footnote{\url{https://speechresearch.github.io/prompttts/}}.

\end{abstract}
\vspace{-0.1in}
\section{INTRODUCTION}
\label{sec:introduction}

Recent research has achieved great success in text and image generation guided with a text description as prompt~\cite{PanGuLA, CPM-2, GAN-CLS, iGPT} (e.g., GPT-3~\cite{GPT-3} or DALLE-2~\cite{Dalle-2}). Beyond text and image generation, there is little research on prompt-based guidance for text-to-speech (TTS) synthesis~\cite{Survey, NaturalSpeech} with different styles such as pitch, speaking speed, and emotion. Previous works on controllable TTS focus on controlling specific style factors: prosody control with word-level prosody tags~\cite{Prosody}, speaking speed control with sentence-level speaking-rate \cite{SCTTS}, and pitch control with pitch contours~\cite{FastPitch}. Style control can be achieved in an explicit manner by using the value of style factors such as pitch~\cite{FastSpeech-2} or an implicit manner by learning a style token~\cite{GST} from the reference speech. However, all previous works require users to provide the specific value of style factors with acoustic knowledge or choose the reference speech that meets the requirements, which are time-consuming and not user-friendly. Therefore, it is a better choice if style control is achieved with a text description in natural language.

In this work, we explore the possibility of leveraging a text description (denoted as \textbf{prompt}) to guide speech synthesis. To be more specific, as shown in Table~\ref{input}, the input prompt consists of a style description (denoted as \textbf{style prompt}) and a content description containing the text to be converted to speech (denoted as \textbf{content prompt}) with a colon in between. For example, an input prompt, ``\textit{A lady whispers to her friend slowly: everything will go fine, right?}'', means that the model needs to synthesize the speech with content of ``everything will go fine, right?'' in a female voice, a slow speaking speed, and a whispering manner. In this way, users are able to create speech from a prompt, resulting in style control without the requirements for acoustic knowledge or reference speech. 

\begin{table*}[!h]\footnotesize
    \caption{The example of prompts containing style prompts and content prompts with colons in between.}
    \label{input}
    \centering
    \begin{tabular}{rl}
        \toprule
        \makebox[0.47\textwidth][c]{\textbf{Style Prompt}} & \makebox[0.47\textwidth][c]{\textbf{Content Prompt}} \\
        \midrule
        \midrule
        I need a man with a deep and loud voice to talk cheerfully: & No important letter come in a parcel, is there? \\
        Please ask a girl to shout loudly with a sharp voice: & But folks hereabouts don't like him. \\
        A lady whispers to her friend slowly: & Everything will go fine, right? \\
        \bottomrule 
    \end{tabular}
    \vskip -0.16in
\end{table*}

\begin{figure*}[!htb]
	\centering
	\vskip 0.02in
    \includegraphics[width=0.50\textwidth]{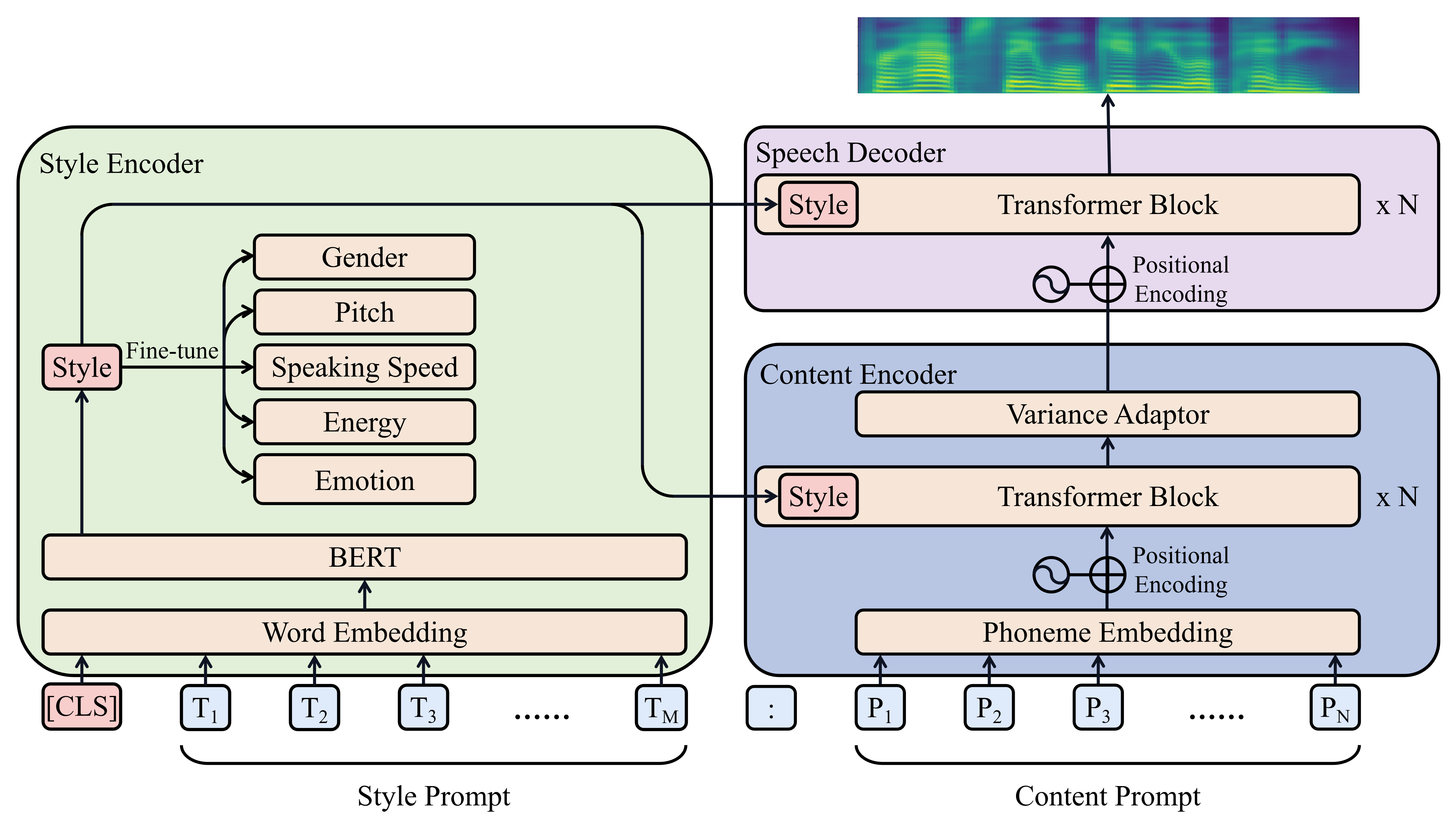}
	\vskip -0.08in
	\captionof{figure}{The architecture of PromptTTS consisting of a style encoder, a content encoder, and a speech decoder.}
	\vskip -0.18in
	\label{model}
\end{figure*}

As the first exploration on TTS guided with prompts, there not exist datasets or systems for this task. Thus, we design a dataset, a system, and an evaluation metric for this task. (1) Dataset: we construct and release a dataset containing prompts with style and content information and the corresponding speech. The prompts describe the speech in 5 style factors including gender, pitch, speaking speed, volume, and emotion. (2) System: to synthesize speech according to a prompt, we propose PromptTTS to serve as baseline for future research in this task, which consists of a style encoder, a content encoder, and a speech decoder. The style and content encoders extract style and content representations from the prompt, respectively. The speech decoder utilizes both representations to synthesize speech accordingly. (3) Evaluation metric: we calculate the accuracy between the style factors from output speech and those from prompts as evaluation metric for style control.

The contributions of our work include: (1) we propose PromptTTS to synthesize the speech that is consistent with prompts in style and content, which is more user-friendly than previous works; (2) we collect and release a dataset consisting of prompts and the corresponding speech for this task; (3) the experiments show that PromptTTS can generate speech with precise style control and high speech quality.

\vspace{-0.1in}
\section{METHOD}
\label{sec:method}

\vspace{-0.1in}
\subsection{Model Overview}
\label{ssec:model overview}

As shown in Fig.~\ref{model}, PromptTTS consists of a style encoder, a content encoder, and a speech decoder. The style encoder maps a style prompt to a semantic space to extract the style representation, which is used to guide the content encoder and the speech decoder. The content encoder takes a content prompt as input to extract the content representation. And the speech decoder concatenates the style representation and the content representation as input to produce the speech that is consistent with both style and content prompts. We introduce the above modules as follows.

\vspace{-0.1in}
\subsection{Style Encoder}
\label{ssec:style}

The style encoder extracts the style representation from the style prompt with a BERT model~\cite{BERT}, which serves as  guidance to control the style of output speech. The input (style prompt) sequence $T = [T_1, T_2, \cdots, T_{M}]$ is prepended with a $[CLS]$ token, converted into a word embedding, and fed into the BERT model, where $M$ refers to the length of style prompt. The hidden vector corresponding to the $[CLS]$ token is regarded as the style representation to guide the content encoder and the speech decoder. In order to better recognize the semantic information related to speech style, the BERT model is fine-tuned on an auxiliary classification task to predict gender, pitch, speaking speed, volume, and emotion information from style prompts. In this way, the BERT model focuses more on style information and thus is more suitable for PromptTTS to guide speech synthesis.

\vspace{-0.1in}
\subsection{Content Encoder}
\label{ssec:content encoder}

The content encoder extracts the content representation conditioned on the style representation from the style encoder, since the variance adaptor~\cite{FastSpeech-2} in the content encoder predicts information such as duration and pitch that is closely related to the style of output speech. We utilize a grapheme-to-phoneme conversion tool~\cite{G2P} to convert the input (content prompt) sequence to a phoneme sequence $P = [P_1, P_2, \cdots, P_N]$, where $N$ refers to the length of phoneme sequence, and we map it to a phoneme embedding to feed into Transformer blocks. Considering that solely adding the style representation to the input embedding may not be powerful enough to guide the model~\cite{P-tuning-v2}, we prepend the style representation to the input of every Transformer block. Following FastSpeech 2~\cite{FastSpeech-2}, the topmost module of the content encoder is a variance adaptor to predict the duration, pitch, and energy, which can provide enough information to synthesize variant speech and alleviate the one-to-many mapping problem in TTS.

\vspace{-0.1in}
\subsection{Speech Decoder}
\label{ssec:speech decoder}

The speech decoder utilizes the style and content representations from both encoders to generate the mel-spectrogram in the corresponding style and content. To be more specific, both representations are concatenated to form the input of the speech decoder. The style representation is also prepended to the input of every Transformer block following the same mechanism in the content encoder.

\vspace{-0.1in}
\section{DATASET}
\label{sec:dataset}

Given that there are no TTS datasets with prompts, we construct and release a dataset called PromptSpeech which consists of speech and the corresponding prompts. We synthesize speech with 5 different style factors (gender, pitch, speaking speed, volume, and emotion) from a commercial TTS API\footnote{\url{https://azure.microsoft.com/en-us/services/cognitive-services/text-to-speech/\#overview}}. The emotion factor has 5 categories\footnote{General, shout, whisper, cheerful, and sad.} and the gender factor has 2 categories. For the rest of style factors including pitch, speaking speed, and volume, we extract the value of style factors from speech with signal processing tools and divide speech into 3 categories (high/normal/low) according to the proportion. Considering that the style of speech can be described in natural language, we ask experts to write style prompts for each category. To further augment the style prompts, we utilize SimBERT~\cite{SimBERT} to generate more style prompts with similar semantics. The speech and the corresponding prompt describing the same style and content are used as paired data in our dataset.

Since there could be a gap between the synthesized speech and the real speech, besides the dataset mentioned above (denoted as \textbf{synthesized} version), we also construct a dataset (denoted as \textbf{real} version) with real speech from LibriTTS~\cite{LibriTTS} by similar construction process. Due to the lack of emotion in LibriTTS, the real version of PromptSpeech only has 4 style factors. As shown in Table~\ref{dataset}, we split both versions of PromptSpeech into training sets and test sets, which are released in public\footnote{\url{https://speechresearch.github.io/prompttts/}}. The experiments are conducted on PromptSpeech to verify the effectiveness of PromptTTS. 

\begin{table}[!h]\footnotesize
    \caption{The number of samples in both versions of PromptSpeech.
    }
    \label{dataset}
    \centering
    \begin{tabular}{ccc}
        \toprule
        \textbf{Version} & \textbf{Training} & \textbf{Test} \\
        \midrule
        \midrule
        Synthesized & 155092 & 5032 \\
        \midrule
        Real & 26588 & 1305\\
        \bottomrule 
    \end{tabular}
    \vskip -0.16in
\end{table}

\vspace{-0.1in}
\section{EXPERIMENT}
\label{experiment}

\vspace{-0.1in}
\subsection{Model Configuration}
\label{ssec:model configuration}

\noindent\textbf{Style Encoder} We utilize a pre-trained BERT model consisting of 12 hidden layers with 110M parameters~\cite{BERT}. The BERT model is fine-tuned on an auxiliary classification task on 5 style factors with P-tuning v2~\cite{P-tuning-v2}.

\noindent\textbf{Content Encoder} The content encoder consists of a variance adaptor and 4 Transformer blocks, where the dimensions of both style and content representations are set to 256. The variance adaptor consists of a duration predictor, a pitch predictor and an energy predictor following FastSpeech 2~\cite{FastSpeech-2}. 

\noindent\textbf{Speech Decoder} The Transformer blocks in the speech decoder and the content encoder share the same model architecture. The output mel-spectrogram of PromptTTS is transformed into speech using a pre-trained HiFiGAN~\cite{HiFiGAN}\footnote{\url{https://github.com/jik876/hifi-gan}}.

\vspace{-0.1in}
\subsection{Baseline System}
\label{ssec:baseline system}

Given that there are no previous works designed for this task, we construct a straight-forward two-stage system as baseline. In the first stage, the system explicitly predicts the value of style factors in style prompts with a BERT-based model fine-tuned with P-tuning v2~\cite{P-tuning-v2}. In the second stage, the value of style factors is converted to a style embedding to guide the style of the output speech~\cite{FastSpeech-2}.

\vspace{-0.1in}
\subsection{Evaluation Metric}
\label{ssec:evaluation metric}

The evaluation metric is the accuracy between the style
factors from output speech and those from prompts. The accuracy of pitch, speaking speed, and volume can be calculated with tools in signal processing\footnote{\url{https://github.com/JeremyCCHsu/Python-Wrapper-for-World-Vocoder}}. For the rest of style factors including gender and emotion that cannot be recognized with signal processing tools, we train classifiers to recognize the categories of output speech. According to Table~\ref{classifier}, the accuracy of classifiers is high enough ($\ge98\%$) to show that: (1) PromptSpeech is well-established and distinguishable; (2) the classifiers can serve as evaluation metric for PromptTTS.

\begin{table}[!h]\footnotesize
    \caption{The accuracy (\%) of classifiers on gender and emotion.}
    \label{classifier}
    \centering
    \begin{tabular}{ccc}
        \toprule
        \textbf{Version} & \textbf{Gender} & \textbf{Emotion} \\
        \midrule
        \midrule
        Synthesized & 99.97 & 98.80 \\
        \midrule
        Real & 98.26 & -\\
        \bottomrule 
    \end{tabular}
    \vskip -0.16in
\end{table}

\vspace{-0.1in}
\subsection{Results}
\label{sssec:results}

In this section, we first introduce the accuracy of PromptTTS and the two-stage baseline system in style control. Then, we conduct ablation studies on the style encoder and analyse the advantages of PromptTTS over the two-stage baseline system. Finally, we conduct mean opinion score (MOS)~\cite{MOS} to evaluate the speech quality from human perspective.

\vspace{-0.1in}
\subsubsection{Main Results}
\label{sssec:main results}

Table~\ref{main result} shows the accuracy of PromptTTS and the two-stage baseline system on 5 style factors. We have the following observations: (1) PromptTTS outperforms the two-stage baseline system on all style factors on PromptSpeech since it could alleviate the cascaded error of the two-stage system, which is discussed in detail in \ref{sssec:performance analysis}; (2) PromptTTS achieves the average accuracy of 90.31\% on PromptSpeech, outperforming baseline by 3.25\%, which shows that PromptTTS can synthesize speech in a more consistent style with the intention of style prompts.

\begin{table}[!h]\footnotesize
    \caption{The accuracy (\%) of PromptTTS and the two-stage baseline system on 5 style factors. Note that emotion is not available in real speech.}
    \label{main result}
    \centering
    \setlength{\tabcolsep}{0.65mm}{
        \begin{tabular}{cccccccc}
            \toprule
            \textbf{Version} & \textbf{Setting} &\textbf{Gender} & \textbf{Pitch} & \textbf{Speed}  & \textbf{Volume} & \textbf{Emotion}  & \textbf{Mean} \\
            \midrule
            \midrule
            \multirow{2}*{Synthesized} & Two-stage & 97.84 & 73.58 & 82.79 & 81.92 & 88.16 & 84.86 \\
            ~ & PromptTTS & 99.57 & 82.60 & 86.55 & 87.35 & 91.78 & 89.57 \\
            \midrule
            \multirow{2}*{Real} & Two-stage & 97.36 & 81.14 & 89.05 & 89.39 & - & 89.24 \\
            ~ & PromptTTS & 99.16 & 82.69 & 92.57 & 89.73 & - & 91.04 \\
            \bottomrule 
        \end{tabular}
    }
    \vskip -0.16in
\end{table}

\vspace{-0.1in}
\subsubsection{Ablation Studies}
\label{sssec:ablation studies}

Since the style encoder plays an important role in learning and controlling the style of output speech, we conduct ablation studies on it by changing fine-tuning methods from P-tuning v2~\cite{P-tuning-v2} to the standard fine-tuning method in the BERT model~\cite{BERT} or no fine-tuning methods at all. From the results in Table~\ref{ablation study}, we can draw the following conclusions: (1) P-tuning v2 slightly outperforms the standard fine-tuning method in BERT model, which proves that P-tuning v2 is suitable for PromptTTS to understand style prompts in PromptSpeech; (2) the large performance drops in the setting without fine-tuning methods indicate that fine-tuning is necessary to help PromptTTS focus on keywords containing style information.

\begin{table}[!h]\footnotesize
    \caption{The accuracy (\%) of PromptTTS with different fine-tuning methods (``FT Method'' in table) in the style encoder. ``Standard'' stands for the standard fine-tuning method in BERT.}
    \label{ablation study}
    \centering
    \setlength{\tabcolsep}{0.45mm}{
        \begin{tabular}{cccccccc}
            \toprule
             \textbf{Version} & \textbf{FT Method} & \textbf{Gender} & \textbf{Pitch} & \textbf{Speed}  & \textbf{Volume} & \textbf{Emotion}  & \textbf{Mean} \\
            \midrule
            \midrule
            \multirow{3}*{Synthesized} & P-tuning V2 & 99.57 & 82.60 & 86.55 & 87.35 & 91.78 & 89.57 \\
            ~ & Standard & 99.02 & 83.40 & 86.27 & 86.39 & 91.40 & 89.30 \\
            ~ & No fine-tuning & 38.21 & 42.45 & 65.37 & 72.25 & 15.94 & 46.84 \\
            \midrule
            \multirow{3}*{Real} & P-tuning V2 & 99.16 & 82.69 & 92.57 & 89.73 & - & 91.04 \\
            ~ & Standard & 99.16 & 81.70 & 91.66 & 89.21 & - & 90.43 \\
            ~ & No fine-tuning & 52.60 & 33.57 & 61.50 & 60.14 & - & 51.95 \\
            \bottomrule 
        \end{tabular}
    }
    \vskip -0.16in
\end{table}

\vspace{-0.1in}
\subsubsection{Performance Analysis}
\label{sssec:performance analysis}

To analyze the advantages of PromptTTS over the two-stage baseline system, we show the average accuracy of the two-stage baseline system at each stage in Table~\ref{performance analysis}. Note that stage one corresponds to the explicit classification on style factors and stage two corresponds to the speech synthesis given ground-truth style factors. From the table, we can see that the average accuracy of each stage is 94.97\% and 91.60\%, respectively, which would propagate mistakes in every stage. However, PromptTTS is an end-to-end system that relies on a latent space to transfer information without cascaded error, thus making it achieve better accuracy.

\begin{table}[!h]\footnotesize
    \caption{The accuracy (\%) of the two-stage baseline system in each stage (``Stage'' in table). Stage one corresponds to the explicit classification on style factors and stage two corresponds to the speech synthesis given ground-truth style factors.}
    \label{performance analysis}
    \centering
    \setlength{\tabcolsep}{1.07mm}{
        \begin{tabular}{cccccccc}
            \toprule
             \textbf{Version} & \textbf{Stage} & \textbf{Gender} & \textbf{Pitch} & \textbf{Speed}  & \textbf{Volume} & \textbf{Emotion}  & \textbf{Mean}\\
            \midrule
            \midrule
            \multirow{2}*{Synthesized} & One & 99.15 & 80.54 & 89.17 & 91.14 & 92.31 & 90.46 \\
            ~ & Two & 98.68 & 91.36 & 92.85 & 89.89 & 95.51 & 93.66 \\
            \midrule
            \multirow{2}*{Real} & One & 100.00 & 84.51 & 96.85 & 89.58 & - & 92.74  \\
            ~ & Two & 97.36 & 96.01 & 91.95 & 99.79 & - & 96.28 \\
            \bottomrule 
        \end{tabular}
    }
    \vskip -0.16in
\end{table}

\vspace{-0.1in}
\subsubsection{Speech Quality}
\label{sssec:speech quality}

To evaluate the perceptual quality, we perform MOS~\cite{MOS} on the test sets. We compare the MOS of audio samples including: (1) GT, the ground-truth recordings; (2) GT mel + HiFiGAN, where we first convert ground-truth speech into mel-spectrogram, and then convert the mel-spectrogram back to speech using HiFiGAN \cite{HiFiGAN}; (3) Two-stage; (4) PromptTTS. Both systems in (3) and (4) use HiFiGAN as vocoder. We also conduct comparison MOS (CMOS)~\cite{CMOS} to compare the speech quality between PromptTTS and the two-stage baseline system. According to Table~\ref{speech quality}, it can be seen that PromptTTS outperforms the two-stage baseline system slightly in terms of speech quality, which is acceptable since the PromptTTS is designed for style control with prompts.

\begin{table}[!h]\footnotesize
    \newcommand{\tabincell}[2]{\begin{tabular}{@{}#1@{}}#2\end{tabular}}
    \caption{The results of speech quality with 95\% confidence intervals.}
    \label{speech quality}
    \centering
    \setlength{\tabcolsep}{2.04mm}{
        \begin{tabular}{cccc}
            \toprule
             \textbf{Version} & \textbf{Setting} & \textbf{MOS} & \tabincell{c}{\textbf{CMOS} \\ (vs. Two-stage)}\\
            \midrule
            \midrule
            \multirow{4}*{Synthesized} & GT & 4.28 $\pm$ 0.10 & - \\
            ~ & GT mel + HiFiGAN & 4.22 $\pm$ 0.07 & - \\
            ~ & Two-stage & 3.95 $\pm$ 0.07 & 0 \\
            ~ & PromptTTS & 4.00 $\pm$ 0.08 & 0.057 \\
            \midrule
            \multirow{4}*{Real} & GT & 4.42 $\pm$ 0.11 & - \\
            ~ & GT mel + HiFiGAN & 4.40 $\pm$ 0.08 & - \\
            ~ & Two-stage & 3.78 $\pm$ 0.08 &  0 \\
            ~ & PromptTTS & 3.80 $\pm$ 0.07 & 0.025 \\
            \bottomrule 
        \end{tabular}
    }
    \vskip -0.16in
\end{table}

\vspace{-0.1in}
\section{CONCLUSION}
\label{sec:conclusion}

In this work, to explore the possibility of guiding TTS with a prompt, we propose PromptTTS that can synthesize the speech consistent with the prompt in style and content. Compared with previous works in controllable TTS, PromptTTS controls the generation of speech in a more user-friendly way. We also collect and release a dataset called PromptSpeech that consists of prompts and the corresponding speech for this task. The experiments verify that PromptTTS can generate speech with precise style control and high speech quality. In the future, we will explore the zero-shot ability of PromptTTS to control unseen style factors and improve the pre-training methods to extract better representations from prompts.

\bibliographystyle{IEEEbib}
\bibliography{refs}

\end{document}